\documentclass[useAMS,usenatbib]{mn2e}

\usepackage{times}
\usepackage{amssymb}
\usepackage{pslatex}

\title[Instability and  Lyapunov times for the 3-body
problem]{On the
relationship between instability and  Lyapunov times for the 3-body
problem}

\author[D. J. Urminsky and D. C. Heggie]{D. J. Urminsky$^{1}$\thanks{E-mail:
david.urminsky@ed.ac.uk} and D. C. Heggie$^{1}$\\
$^{1}$ School of Mathematics and Maxwell Institute for Mathematical
  Sciences, University of Edinburgh, \\
James Clerk Maxwell Building, The
  King's Buildings, Edinburgh, United Kingdom, EH9 3JZ}

 \usepackage{graphicx}

\begin{document}

\pagerange{\pageref{firstpage}--\pageref{lastpage}} \pubyear{2008}

\maketitle

\label{firstpage}

\begin{abstract}
In this study we consider the relationship between the survival time  and
the Lyapunov time
for 3-body systems.  It is shown that the
Sitnikov problem exhibits a two-part power law relationship as
demonstrated in \cite{mikkola2007} for the general 3-body problem.  
Using an approximate Poincar\'e
map on an appropriate surface of section, we delineate escape
regions in a domain of initial conditions and use these regions to
analytically obtain a new functional relationship between the Lyapunov
time and the survival time for the 3-body problem.  The marginal
probability distributions of the Lyapunov and survival times are
discussed and we show that the  probability density function of
Lyapunov times for the Sitnikov problem is  similar to that for the
general 3-body problem.
\end{abstract}


\section{Introduction}

A correlation between the Lyapunov time,  the time it takes for nearby
orbits to diverge by $e$,  and the time in
which an orbit undergoes a sudden transition in the Solar
System was first discussed by  \cite{lecar1992}.  From a
study of orbits of asteroids \citep{soper} between Jupiter
and Saturn, the authors noted a relationship between the
Lyapunov time, $t_l$, and the time, $t_d$, which an asteroid takes
to cross the orbit of Jupiter or Saturn.   A correlation between
these two time scales was also
noted for asteroids in the outer asteroid belt in
\cite{lecar1992b}.  In both studies it was found that
$t_d$ and $t_l$ are related by
\begin{equation}\label{eq:pl}
  \frac{t_d}{C} = A\left(\frac{t_l}{C}\right)^\beta,
\end{equation}
where $\beta$ is a constant,  $C$ is a normalization constant,
and $A$ is a constant of proportionality.  

In support of the relationship (\ref{eq:pl}), 
\cite{lecar1992} considered the elliptic restricted 3-body
problem  in which the massless particle, $m_3$, began its
motion around the secondary mass which was $1/9$ the mass
of the primary where the orbit of the secondary body had
an eccentricity of
0.1.  In this example, $t_d$ was taken to be the time it
took for $m_3$ to escape via one of the collinear
Lagrange points.  Correlating data from 1000 orbits, the
study found that  (\ref{eq:pl}) holds for
$\beta \approx 1.8$.

There have been many other investigations into the relationship
between Lyapunov times and survival times. 
\cite{levison1993} in a study of  Edgeworth-Kuiper belt
objects showed a relationship between the Lyapunov time and
the time it takes for these object to cross the orbit of
Neptune.    In this study, the authors considered orbits
of 200 particles with eccentricities between 0.01 and 0.1.
They found that  (\ref{eq:pl}) holds but
with a slightly higher exponent value $\beta\approx 1.9$. 
In another study,  \cite{murison1994} considered the
restricted elliptic 3-body problem with Jupiter as the
secondary mass.  Again, they found that the relationship
(\ref{eq:pl}) holds with $\beta= 1.74 \pm 0.03$.
 
Despite all the support of the relationship (\ref{eq:pl}),
there have been some disagreements with the relationship.
\cite{murray1997} found that the relationship does not
hold for some bodies in the outer belt.  Their explanation
is based on the properties of a system controlled by a 
critical KAM curve. The dynamics of the outer asteroid belt
are not controlled by a single critical KAM curve, and
the authors argued that this means that there is no reason
to expect a simple scaling between the Lyapunov time and
the escape time.  In another study  \cite{morbidelli1996}
gave a two part relationship. They suggested that for orbits in the
Nekhoroshev regime, the relationship between $t_d$ and
$t_l$ should be exponential (i.e. $t_d \sim \exp(t_l)$)
whereas in a regime with resonance overlapping a
relationship of the form  (\ref{eq:pl}) can hold.  
 
 An investigation  of the relationship
between escape times and Lyapunov times for the general
3-body problem has recently been conducted by 
\cite{mikkola2007}.  The authors looked for a correlation
between the Lyapunov time and escape time for the planar
3-body problem.  The authors considered over 10000 initial
values in a domain of initial conditions and found that
a two part power law works best.  For orbits with small $t_l$, the
authors suggested that a power law  (\ref{eq:pl}) with exponent
$\beta\approx 2.3$ approximates the data whereas for  large $t_l$,
the power law  fits better with  $\beta\approx 1$.   

In this study we discuss the relationship between the Lyapunov time
and the survival time for a specific 3-body configuration known as the
Sitnikov problem.  In section \ref{sec:sitode}  we
demonstrate that $t_l$ and $t_d$ for orbits of the Sitnikov problem
exhibit a two part power law relationship similar to that for the
general 3-body problem. It is further shown that the relationship
between $t_l$ and $t_d$ for small $t_l$ is dependent on the
eccentricity of the binary system.  In section \ref{sec:sitmap} we
present an approximate map for the Poincar\'e map discussed in
\cite{moser} for the Sitnikov problem.  We then
demonstrate that the relationship between $t_l$
and $t_d$, for orbits computed with the approximate map, is similar to
both the Sitnikov problem and the general 3-body problem.
Using the approximate Poincar\'e map we delineate a region of initial
conditions which escape quickly;  orbits for these initial conditions 
are used to construct a new functional relationship between $t_l$ and
$t_d$. Finally, in section \ref{sec:distributions} we discuss the
probability distributions of $t_l$ and $t_d$ and compare them to the
distributions for the general 3-body problem.

\section{Sitnikov Problem}\label{sec:sitode}
\begin{figure}
  \includegraphics[width=84mm]{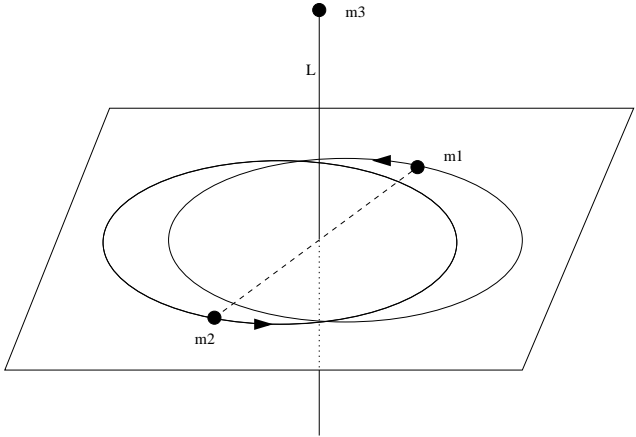}
  \caption{The Sitnikov Problem}
  \label{fig:sitnikov}
\end{figure}

The Sitnikov problem is the problem of the motion of a
 massless particle, $m_3$, on the axis of symmetry, $L$, of
an equal mass ($m_1=m_2$) binary
(Figure \ref{fig:sitnikov}).
 Units are chosen such that the total mass of the binary 
 is unity, the period of the binary is $2\pi$ and the
 gravitational constant $G=1$.   The equation of motion
 for $m_3$ takes the form 
\begin{equation}
\label{eq:sitnikov}
  \ddot{z} = - \frac{z}{\sqrt{z^2+r^2}^3}
\end{equation}
where $z$ is the position of $m_3$ along $L$, $z=0$
 corresponds to the plane of the binary, and $r$ is the
 distance
from one of the binary particles to the centre of mass.
  The specific energy
 for $m_3$ is given by
\begin{equation}\label{eq:energy}
  E=\frac{1}{2}\dot{z}^2 - \frac{1}{\sqrt{z^2+r^2}}.
\end{equation}
 The value of $r$ can be computed from Kepler's equation
 or for small
eccentricities, $\varepsilon$, of the binary, we can approximate $r$
 to first order in $\varepsilon$ by,
\begin{equation}\label{eq:rapprox}
 r \approx \frac{1}{2}\left(1-\varepsilon\cos\left(t\right)\right).
\end{equation}  
 
\subsection{Definitions}\label{sec:def}	
The survival time of orbits for the Sitnikov Problem is defined as the
duration of the numerical experiment to the point where $m_3$ escapes from
the system. We say $m_3$ has escaped at time $t$ if sign($z(t)$) =
 sign($\dot{z}(t)$) and 
\begin{equation}\label{eq:cond1}
K_\lambda=\frac{1}{2}\dot{z}(t)^2 - \frac{1}{\sqrt{z(t)^2+\lambda^2}} > 0.
\end{equation}
where $\lambda$ is a constant such that $\lambda=(1-\varepsilon)/2$.
It can be shown \citep{urminskyPHD} that if the motion of $m_3$
satisfies the above conditions, then the system's final motion is
hyperbolic-elliptic.

The Lyapunov time for orbits of the Sitnikov problem can be computed
from the solutions of the variational equations, $\delta z(t)$.  For
chaotic systems, the magnitude of the variational solutions has order
\begin{equation}\label{eq:varmag}
\frac{|\delta z(t)|}{|\delta z(0)|} \sim \exp\left(\frac{t}{t_l}\right),
\end{equation}
where $t_l$ is the Lyapunov time.
Evaluating (\ref{eq:varmag}) at time $t=t_d$ and solving
for $t_l$ gives the Lyapunov time as
\begin{equation}\label{eq:lyap}
 t_l = \frac{t_d}{\ln\left(\left|\delta z(t_d)\right|/\left|\delta z_0 \right|\right)},
\end{equation}
where $\delta z_0=\delta z(0)$.

\subsection{Initial Conditions}\label{sec:sit_initial}
If we take initial conditions for (\ref{eq:sitnikov}) such
that  $z(t_0)=0$ for an initial time $t_0=0$, we can determine from
(\ref{eq:cond1}) that as $m_3$ crosses the plane of motion
of the binary, a velocity of
\begin{equation}\label{eq:escvel}
  \dot{z}(t_0)>\sqrt{\frac{2}{\lambda}}
\end{equation}
will ensure that $m_3$ escapes the system without returning
to the plane of the binary.  As (\ref{eq:rapprox}) is periodic with
period $2\pi$,  
we can consider initial conditions in polar coordinates where time
is the angular argument and $\dot{z}(t_0)$ is the radial
argument.  Thus we can define the set of initial conditions for 
(\ref{eq:sitnikov}) as the  circle of  radius $\sqrt{2/\lambda}$ centred 
at the origin.  
\begin{figure}
  \includegraphics[width=\linewidth]{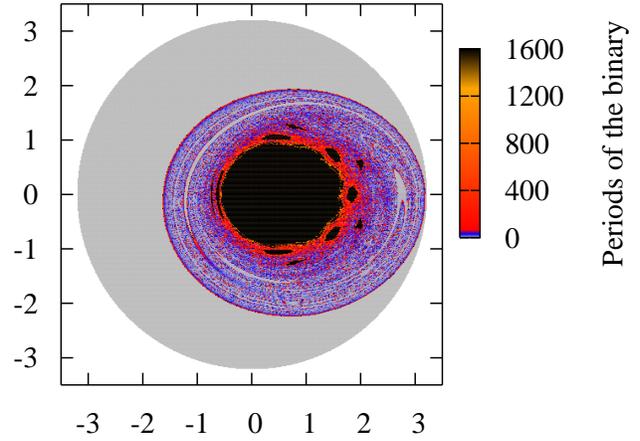}
  \caption{Survival time of orbits of the Sitnikov problem
    with initial conditions in the
    circle whose radius is given by (\ref{eq:escvel}). 
    Each initial condition is
    plotted in polar coordinates where the radial argument is determined
    by the initial velocity and the angular argument is determined by
    $t_0$.  The colour associated with each point indicates the number of
    periods of the binary before escape was determined.}
  \label{fig:fractal}
\end{figure}

Figure \ref{fig:fractal} shows the complement of the region defined by (\ref{eq:escvel}).
  A grid of initial
conditions was chosen inside the disk and iterated forward using the
Bulirsch-Stoer method (\cite{Press}) for either 10000 time units or
until the escape criterion was satisfied.  The colour associated with each
initial condition represents the number of periods of the binary
before $m_3$
either satisfied the escape criterion, or the numerical integration
algorithm reached its maximum time.  The outer grey  region
represents initial conditions in which the escape criterion was
satisfied before the mass returned to the plane of motion of the
binary.   The inner black regions correspond to initial
conditions whose orbits remain bounded. 

\subsection{Results}

To demonstrate a relationship between the survival time and the
Lyapunov time for the Sitnikov problem, 10000 initial conditions
where chosen in the region described in section \ref{sec:sit_initial}.
The Bulirsch-Stoer method \citep{Press}  was chosen as the numerical 
integrator with a relative tolerance of $10^{-12}$.  We
integrated each initial condition simultaneously with the variational
equations for $100000$ time units or until
the solution satisfied the escape criterion.  If the solution
failed to satisfy the escape criterion within the time limit given to the
integrator it was not considered in the results.  This is because
there is a large area of bounded motion, approximately the black
regions in Figure  \ref{fig:fractal}, which  never escape.  

\begin{figure}
  \includegraphics[width=84mm]{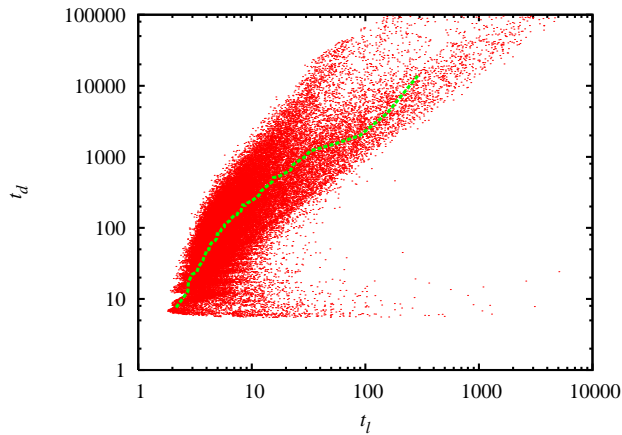}
  \caption{The scatter diagram of the survival time, $t_d$, and the
    Lyapunov time, $t_l$, for the Sitnikov problem where the
    eccentricity of the binary is $\varepsilon=0.61$.  The dashed line
    is a median curve such that at any position along  the curve there
    are an equal number of scatter points above and below the line.}
  \label{fig:scatter1}
\end{figure}

\begin{figure}
  \includegraphics[width=84mm]{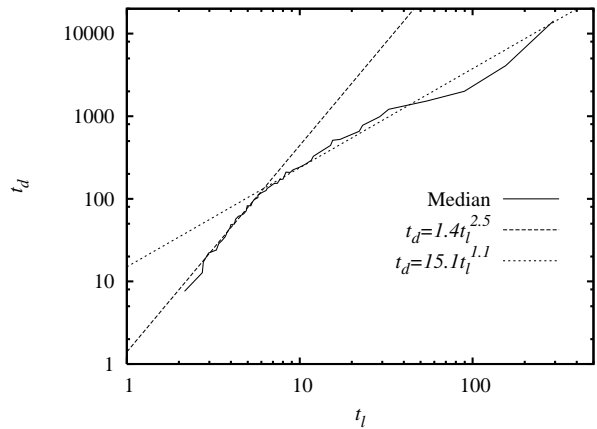}
  \caption{The solid curve is the median curve shown in Figure
  \ref{fig:scatter1}. The two dashed lines represent the power law
  relationship (\ref{eq:pl}) on different time intervals.  For
  $1<t_l<6)$ we find that $\beta\approx 2.5$ best approximates  the
  data. For $t_l>6$ we find that $\beta\approx 1.1$ works better. }
  \label{fig:ode2part}
\end{figure}

Figure \ref{fig:scatter1} displays the  $(t_l,t_d)$ scatter diagram in
 logarithmic scale for $\varepsilon=0.61$. As in 
\cite{mikkola2007}, the range of $t_l$ is
divided into 50 intervals each containing an equal number of points.
The dashed curve in Figure \ref{fig:scatter1} displays the median of the
survival time, $t_d$, in each of the 50 intervals. Comparing this plot
to Figure 3 in  \cite{mikkola2007} we note some similarities.  First,
 for small $t_l$ the median curve is steeper compared to larger $t_l$
 values.  Secondly, the density of the scatter points becomes smaller
 as $t_l$ increases.  Finally, for small $t_d$ the scatter plot has
 horizontal band-like structures.
In Figure \ref{fig:ode2part} we re-plot the median curve in Figure
\ref{fig:scatter1} and approximate the median curve with (\ref{eq:pl})
 on two separate $t_l$ intervals.  For $1<t_l<6$ we find that
 (\ref{eq:pl})
approximates the curve with $\beta=2.5$.  For $t_l>6$ we find that
(\ref{eq:pl}) approximates the data with $\beta=1.1$.  

\subsection{Perturbations caused by large eccentricities}

Of the systems studied by \cite{mikkola2007}, in particular the
free-fall 3-body problem, the change in energy of the escaping body
can vary widely depending on the  interaction with the resulting binary
system. 
  This can be modelled in the Sitnikov problem by
increasing the eccentricity of the binary. 

\begin{figure}
  \includegraphics[width=84mm]{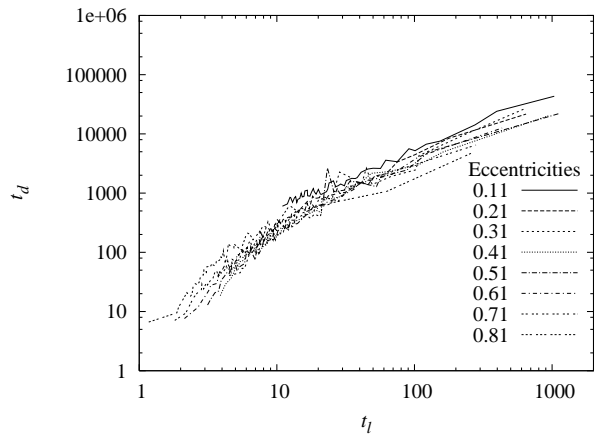}
  \caption{The median $(t_l,t_d)$ curve for varying eccentricities  of
  the Sitnikov problem. }
  \label{fig:sitecc}
\end{figure}

Figure \ref{fig:sitecc} represents the median curves for a series of
experiments in which the eccentricity of the binary system in the
Sitnikov problem is varied.   One distinguishing feature in this
figure is the increasing prominence  of a two part power law
relationship between the  survival time and the Lyapunov time as the
eccentricity increases.   Interestingly, the minimum Lyapunov and
survival times also decrease  as the eccentricity increases.  

In summary, by increasing the eccentricity of the Sitnikov problem we  can
cause large perturbations to the energy of $m_3$ and for $\varepsilon$
large enough the two-part power law becomes more prominent.  An
explanation for the different power laws between small and large $t_l$
is still needed.  To help provide further theoretical explanations for the
relationship we can turn to an approximate Poincar\'e map for the
Sitnikov problem derived in \cite{urminskyPHD}. 

\section{Approximate Poincar\'e map}\label{sec:sitmap}

The plane which corresponds to $z=0$ is a natural choice for a
surface of section (SOS)
on which to study escape with the Sitnikov problem.   On the SOS we can
consider a map $\phi:(v_0,t_0)\rightarrow(v_1,t_1)$
which takes $m_3$ from one crossing of the SOS to the next
crossing.  If $m_3$ is on the SOS at time $t_0$, $\phi$ is a map which
brings $v_0=\dot{z}(t_0)$ to time $t_1>t_0$ where
$v_1=\dot{z}(t_1)$ and $z(t_1)=0$.  \cite{moser} shows that there
exists a real analytic simple closed curve in $\mathbb{R}^2$ in whose
interior, $D_0$, the mapping $\phi$ is defined.  In addition, $\phi$
maps $D_0$ onto a domain $D_1$ and for $\epsilon>0$ the boundary curves for
$D_0$ and $D_1$ intersect transversally. Any point not in $D_0$ is
said to escape.

Capturing the dynamics of the map $\phi$ can provide insights into the relationship
between $t_l$ and $t_d$. To do this, we  consider the following
symplectic map which approximates  $\phi$ \citep{urminskyPHD},
including the
approximation (\ref{eq:rapprox}), from one crossing of the SOS at time
$t_0$ to the next crossing at time $t_1$ given by $\Phi:(t_0,E_0)\rightarrow (t_1,E_1)$, where
\begin{equation}\label{eq:map}
\begin{array}{lcl}
  E_{1/2} & = & E_0 + a \cos\left(t_0\right) + b
  \sin\left(t_0\right) \\
  t_{1/2} & = & t_0 + \alpha\left(-E_{1/2}\right)^{-3/2}  \\
  t_{1} & = & t_{1/2} + \alpha\left(-E_{1/2}\right)^{-3/2}  \\
  E_{1} & = & E_{1/2} - a \cos\left(t_1\right) + b
  \sin\left(t_1\right), 
\end{array}
\end{equation}
and $a$, $b$ and $\alpha$ are constants.  The quantities $t_{1/2}$
and $E_{1/2}$ are the time and energy values of $m_3$,
respectively, when $m_3$ reaches a local maximum  distance from the SOS with
$\dot{z}(t_{1/2})=0$. The map is derived by approximating the change
in energy of $m_3$
  on two  
time intervals in which we approximate its orbit by an orbit which
escapes parabolically.
  The first
time interval  $(t_{0},t_{1/2})$ corresponds to $m_3$ 
 moving  away from the
SOS.  The second time interval   $(t_{1/2},t_{1})$ corresponds to the
period in which  $m_3$ 
returns to the SOS.
It is clear that the change in energy is periodic in $t_0$ and the
trigonometric terms in (\ref{eq:map}) can be thought of as a lowest
order Fourier approximation to this change.
The change in time is approximated by Keplerian
motion over each time interval which means, for the chosen units, $\alpha=\pi/(2\sqrt{2})$.   The map can be generalized as the
iterative map $\Phi:(t_n,E_n)\rightarrow (t_{n+1},E_{n+1})$
and the constants $a$ and $b$
are approximately proportional to $\varepsilon$ with 
 \begin{equation}
	\begin{array}{rcl}
		a & \approx & 0.599 \;\;\varepsilon/4 \\
		b & \approx & 2.029 \;\;\varepsilon/4.
	\end{array}
\end{equation}
Sometimes it is more useful to write (\ref{eq:map}) in the form
\begin{equation}\label{eq:mapleap}
	\begin{array}{rcl}
	t_n & = & t_{n-1} + 2\alpha(-X_{n-1})^{-3/2} \\
	X_{n} & = & X_{n-1} + 2b\sin(t_n)
	\end{array} 
\end{equation}
for $n=1,2,3...$ where $X_0=E_{1/2} = E_0 + a\cos(t_0) + b\sin(t_0)$,
and generally $X_n = E_{n+1/2}$.

\subsection{Initial conditions}
Analogous to Moser's $D_0$ and $D_1$ for (\ref{eq:sitnikov}), we can
define an open domain $U_0$ for which $\Phi$ is defined
which is mapped into an open region $U_1$.
Since time enters into the change in energy  with
period $2\pi$, and we can transform energy values into velocity values
by (\ref{eq:energy}), we can consider $U_0$ in polar
coordinates where the angular argument is determined by $t$ and the
radial argument is determined by $v$.
An upper bound on allowable energy values in $U_0$ is given by
\begin{equation}\label{eq:fboundary} 
E^f(t)=-a\cos\left(t \right)-a\sin\left(t \right),
\end{equation}
for $t \in [0,2\pi]$ which corresponds to $E_{1/2}=0$ for which the
map is undefined.  All points in
\begin{figure}
  \includegraphics[width=\linewidth]{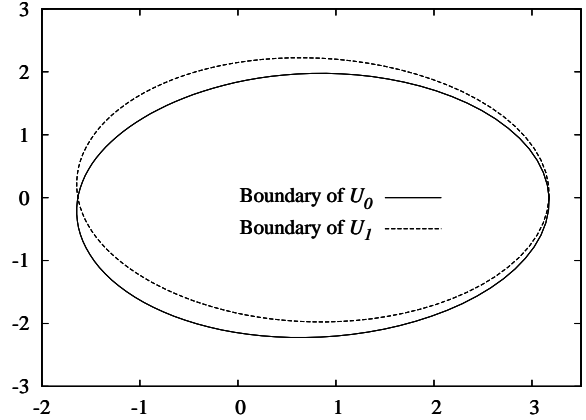}
  \caption{The boundaries $\partial U_0$ and $\partial U_1$ for the
  regions $U0$ and $U_1$ for $\varepsilon = 0.61$.}
  \label{fig:dD0dD1}
\end{figure}
$U_0$ get mapped to the open set $U_1$ whose boundary, 
$\partial U_1$, is defined by
\begin{equation}\label{eq:bboundary}
E^b(t) =
-a\cos\left(t \right)+b\sin\left(t \right),
\end{equation}
for $t \in [0,2\pi]$.  The boundaries $\partial U_0$ and
 $\partial U_1$ are depicted in Figure \ref{fig:dD0dD1}.    
Allowable energy values in $U_0$ at time $t$ 
satisfy $E<E^f(t)$.
To satisfy the physical constraints  of the Sitnikov problem, energy
 values in the domain $U_0$ are also bounded from below.  From 
(\ref{eq:energy}), energy values on the SOS must satisfy 
\begin{equation}\label{eq:Econstraint}
	E+\frac{1}{r(t)} \geq 0.
\end{equation}

It has been shown by \cite{urminskyPHD}, that the dynamics of orbits
with initial conditions in $U_0$ for the map $\Phi$ are
similar to the dynamics of orbits with initial conditions in $D_0$ for the map
$\phi$.  More specifically, it was shown that  the map $\Phi$
satisfies lemmas similar to those proved by \cite{moser} which prove
the existence of a  set $\Lambda\in U_0$ on which the
dynamics are topologically equivalent to the shift map 
on the set of bi-infinite sequences.
   
Initial conditions in $U_0$ can be iterated forwards using
(\ref{eq:map}) until the resulting orbits take on energy and
time values which are outside the domain $U_0$. A comparison
of the Poincar\'e map $\phi$ and the approximate Poincar\'e map $\Phi$
can be found in \cite{urminskypro} in which regions of initial values
on the SOS are delineated by the number of excursions from the SOS an
orbit makes before escaping.

\subsection{Definitions}

 For a given orbit $\mathcal{Z}=\{(t_i,E_i)\}_{i=0}^N$ computed by
 $\Phi$, where $N$ is the number of excursions from the SOS before
 $m_3$ escapes, the survival time is defined
to be $t_d=t_N-t_0$.  The growth of the logarithm of the solutions to 
 the variational equations for the orbit $\mathcal{Z}$ is approximated
 by,
\begin{equation}\label{eq:denom}
\ln|\delta \mathbf{Z}_N| \approx \sum_{i=0}^{N} \ln\left|\mathbf{w}_i\right|, 
\end{equation}
where $\mathbf{Z}_N=(t_N,E_N)$ and $\mathbf{w}_i$ is determined by,
\begin{equation}\label{eq:thevs}
  \mathbf{w}_i = \mathbf{J}_{i-1}\frac{\mathbf{w}_{i-1}}{\left|\mathbf{w}_{i-1}\right|}, \;\;\;\mbox{for}\;\; i=1,...,N,
\end{equation}
in which $|\mathbf{w}_0|= 1$ is chosen at random
and $\mathbf{J}_i$ is the Jacobian of $\Phi$ at time step $i$. After
a few iterations $\mathbf{w}_i$ is aligned with the unstable
direction associated with the solution at the $i$th time step. In a
similar way to equation (\ref{eq:lyap}), we can express the
relationship between $t_d$ and $t_l$ as,
\begin{equation}\label{eq:lyaptimemap}
t_l \approx \frac{t_d}{\displaystyle\sum_{i=0}^{N} \ln\left|\mathbf{w}_i\right| }.
\end{equation}

\subsection{Numerical Results}
\begin{figure}
  \includegraphics[width=\linewidth]{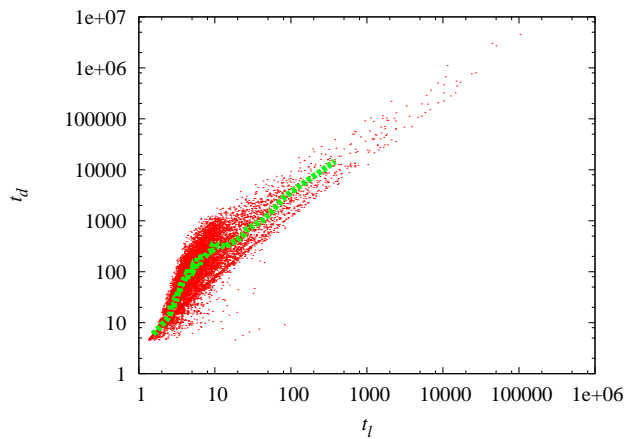}
  \caption{The $(t_l,t_d)$ scatter plot for 10000 uniformly
  distributed initial conditions in the domain of the map $\Phi$ for
  $\varepsilon=0.61$. The dashed line is the median curve associated with the
  scatter plot.}
  \label{fig:scatter2}
\end{figure}

 $10000$ random uniformly distributed initial conditions were
chosen in $U_0$ such that, when iterated using equation
(\ref{eq:map}), they escaped within $1000$ iterations.  If they failed
to escape we did not include them in the calculations as there are regions
of initial conditions 
in $U_0$ for which the corresponding orbits never escape. 
  While
calculating the orbit we simultaneously compute (\ref{eq:denom}) so as to
 determine  the Lyapunov time by equation (\ref{eq:lyaptimemap}).  
Figure \ref{fig:scatter2} shows the $(t_l,t_d)$
scatter plot where the dashed line represents the median curve 
associated with the scatter plot.
\begin{figure}
  \includegraphics[width=\linewidth]{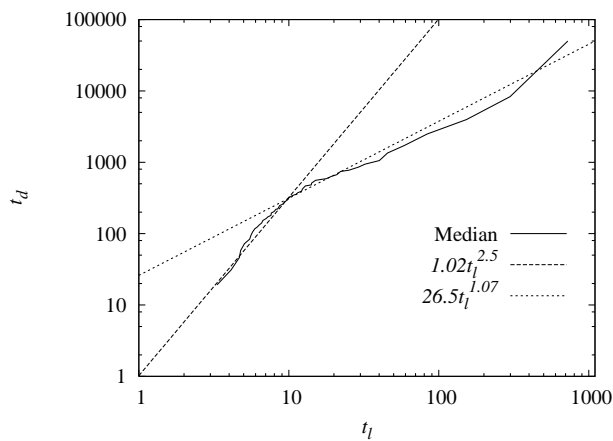}
  \caption{Two part power law relationship
  for 100000 initial conditions
    for the map ($\varepsilon=0.61$)}
  \label{fig:twopartmap}
\end{figure}
Notice that the horizontal spread of the scatter plot for small time
values found in Figure \ref{fig:scatter1} is present in Figure
\ref{fig:scatter2}. In addition, the density of the plotted points
decreases as $t_l$ increases.

\begin{figure}
  \includegraphics[width=\linewidth]{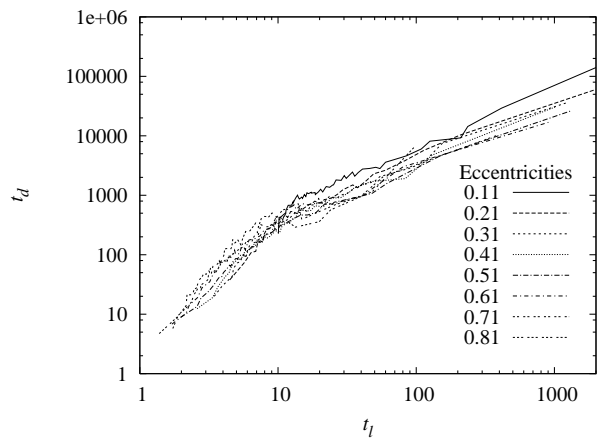}
  \caption{Median curves for the map $\Phi$ for varying eccentricity values. }
  \label{fig:mapvaryecc}
\end{figure}

In Figure \ref{fig:twopartmap} we plot the median curve for the $(t_l,t_d)$
scatter plot for 100000 uniformly distributed initial conditions for
$\varepsilon=0.61$.  Again, there appears to be a
two-part power law relationship for  $t_l$.  The power
law (\ref{eq:pl}) with $\beta=2.5$ approximately fits the median curve on the interval $1<t_l<9$,
whereas a power law with $\beta=1.07$ fits better for $t_l>9$.  Figure
\ref{fig:mapvaryecc} shows the median  curves for various eccentricities
of the binary.  As in Figure \ref{fig:sitecc}, as the eccentricity of
the binary increases, the two-part power law become apparent.

\begin{figure}
  \includegraphics[width=\linewidth]{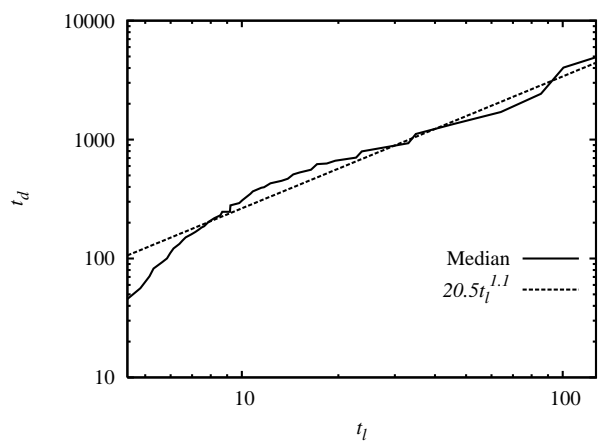}
  \caption{ The median curve on the interval
  containing  90 per cent of the points in Figure
  \ref{fig:twopartmap}.  On this interval a power law relationship
  with $\beta\approx 1.1$ fits the data best.}
  \label{fig:90per}
\end{figure}

Finally, we consider the time interval $4.4<t_l<129.7$ which contains
90 per cent of the scatter points in Figure \ref{fig:scatter2}.  This
interval was chosen such that 5 per cent of the scatter points were in
the region $t_l<4.4$ and 5 per cent of the points were in the region
$t_l>129.7$.  It was found that in this region a power law
relationship with $\beta\approx 1.1$ best fitted the median curve
(Figure 
\ref{fig:90per}).  This
contrasts with the result in \cite{mikkola2007} for the general 3-body
problem where it was found that a power law with $\beta\approx 1.8$
roughly fits 90 per cent of the data on an interval
$.94<t_l<35.2$. Since the data in Figure \ref{fig:90per} is
distributed over a larger interval, the data for larger $t_l$ values
dominates the approximation and the smaller power law approximation
best fits the data.  We shall demonstrate in section
\ref{sec:distributions} that the distribution of $t_d$ for the map is
different than that found for the general 3-body problem which may
account for the discrepancy between the two results.

\subsection{Delineating the region corresponding to escape after one
  excursion}

\begin{figure}
\centering
\includegraphics[width=\linewidth]{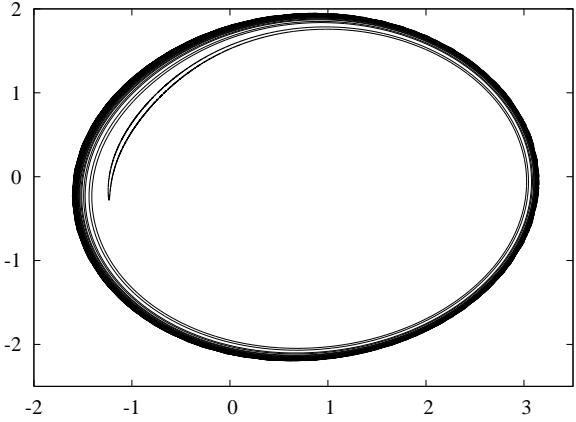}
\caption{The boundary of the region $B_1$.}\label{fig:B1boundary}
\end{figure}

Initial conditions along the boundary $\partial U_0$ defined
by (\ref{eq:fboundary}) lead to  $E_{1/2}=0$.  From equation
(\ref{eq:map}), we can see that these initial conditions lead to a
time $t_{1/2}$ which is undefined. Initial conditions on $E^f(t)$ for
$t \in(0,\pi)$ are contained in the set $D_1$ and so are in the domain
of the inverse map $\Phi^{-1}$.   Iterating these initial conditions
backwards once gives the boundary of the region, $B_1$, corresponding
to  penultimate crossings of the SOS before escape.  
The boundary of $B_1$ can be shown to be given parametrically by,
\begin{equation}\label{eq:B1bound}
\begin{array}{rcl}
 t_{*} & = & t - 2\alpha(2b\sin(t))^{-3/2} \\ 
 E_{*} & = & -2b\sin(t) - a\cos(t_{*}) - b\sin(t_{*}). 
\end{array}
\end{equation}
This boundary is shown in Figure \ref{fig:B1boundary} using 
 polar co-ordinates  where the angular argument is time and the
 radial argument is the velocity of $m_3$ obtained from
 (\ref{eq:energy}).
Iterating initial values on this boundary forward in time using
 (\ref{eq:map}) leads to energy values $E_{3/2}=0$ and from
 equation (\ref{eq:map}) we get,
\begin{equation}\label{eq:e32}
	E_{3/2} = E_{1/2} + 2b\sin(t_1)=0.
\end{equation}
Writing (\ref{eq:e32}) in terms of $E_0$ and $t_0$ and rearranging
 gives,
\begin{equation}\label{eq:e32b}
	\begin{array}{l}
		E_0+a\cos(t_0) + b\sin(t_0) = \\
		\;\;\; -2b \sin\left(t_0 + 2\alpha(-E_0-a\cos(t_0) -b\sin(t_0))^{-3/2}\right).
	\end{array}
\end{equation}
As the boundary of $B_1$ spirals inside $U_0$ (Figure
\ref{fig:B1boundary}) it approaches
the boundary $\partial U_0$. 
 Consider $E_0$ values on the boundary
of $B_1$ for a fixed $t_0$. Define
\begin{equation}\label{eq:theta}
	\theta = 2\alpha\left( -E_0 -a\cos(t_0) -b\sin(t_0)\right)^{-3/2}
\end{equation}
and note that $\theta\rightarrow\infty$ as $E_0\rightarrow E^f(t_0)$. 
Using equation (\ref{eq:theta}) we can re-write (\ref{eq:e32b}) as
\begin{equation}\label{eq:e32c}
	-\left(\frac{\theta}{2\alpha} \right)^{-2/3} = 2b\sin(t_0+\theta),
\end{equation}
and as $\theta\rightarrow\infty$ solutions to (\ref{eq:e32c}) are 
asymptotically approximated by,
\begin{equation}\label{eq:e32d}
	\theta + t_0 \approx k\pi,
\end{equation}
for large $k\in\mathbb{Z}$.  Substituting equation (\ref{eq:theta})
 into (\ref{eq:e32d}) we can obtain the approximation for the energy
 values on the boundary of $B_1$ 
\begin{equation}\label{eq:Eb1}
	E_0^k \approx -\left( \frac{k\pi -t_0}{2\alpha}\right)-a\cos(t_0) -b\sin(t_0),
\end{equation}
for large $k\in\mathbb{Z}$.  Initial conditions on $B_1$ survive only
 one iteration of the map $\Phi$. Substituting (\ref{eq:Eb1}) into
 (\ref{eq:map}) gives the survival time for orbits of initial
 conditions on the boundary of $B_1$ as
\begin{equation}\label{eq:Tb1}
	t_d = t_1 - t_0 \approx 2\alpha \left(
	\left(\frac{k\pi-t_0}{2\alpha}\right)^{-2/3}\right)^{-3/2} =
	k\pi - t_0.  
\end{equation}

\subsection{A functional relationship between $t_l$ and $t_d$}
To derive a functional relationship between $t_l$ and $t_d$, we 
consider initial conditions on the boundary of $B_1$ for a 
fixed $t_0$. For $t_0=\pi/2$, initial conditions on $\partial B_1$ are
 approximated by,
\begin{equation}\label{eq:b1initial}
	\begin{array}{lcl}
		t_0 & = & \pi/2, \\ 
		E_0 & = & - \left(\displaystyle \frac{(k-1/2)\pi}{2\alpha}
\right)^{-2/3} - b,
	\end{array}
\end{equation}
for large $k\in\mathbb{Z}$.  Using the initial conditions
(\ref{eq:b1initial}) and the corresponding survival time given by
(\ref{eq:Tb1}), we can compute the Lyapunov time $t_l$  from
(\ref{eq:lyaptimemap}) for a given initial $|\mathbf{w}_0|=1$.    For
long lived orbits it may be assumed that  $\mathbf{w}_k$ will normally become aligned with
the unstable direction after only a few iterations.  For initial
conditions along $\partial B_1$ on the other hand, this assumption is
invalid since the orbit escapes after only one excursion. The choice
of $\mathbf{w}_0$ has an important
role in determining the growth of $|\delta \mathbf{Z}_1|$ as it
may not necessarily be aligned with the most unstable direction.

\begin{figure}
  \includegraphics[width=\linewidth]{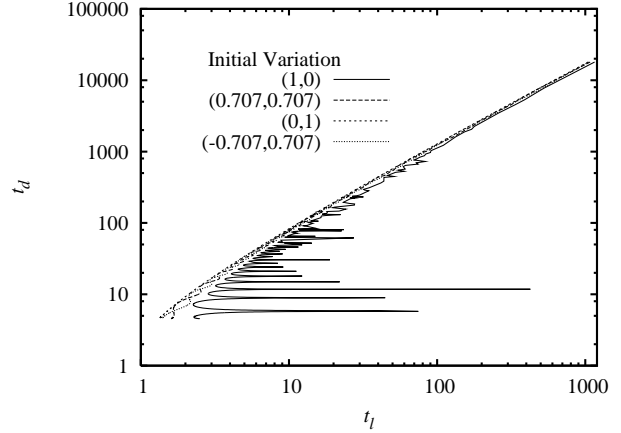}
  \caption{The $(t_l,t_d)$ scatter plot for orbits with initial values on the  boundary of $B_1$ for varying initial vector $\mathbf{w}_0$ (
$\varepsilon=0.61$).  }
  \label{fig:v0vary}
\end{figure}

The vector $\mathbf{w}_0$ should be chosen so as to give the
largest
$|\delta \mathbf{Z}_1|$ possible. Choosing $\mathbf{w}_0$ in this way
has the effect
of making $t_l$ small.  Figure \ref{fig:v0vary} shows the $(t_l,t_d)$
scatter plot for initial values given by (\ref{eq:B1bound}) for various
choices of $\mathbf{w}_0$.  It was found that the vector
$\mathbf{w}_0=[0,1]$ maximizes $\ln|\delta\mathbf{Z}_1|$ and ensures that
$\ln|\delta\mathbf{Z}_1|>0$ for orbits of initial conditions on the boundary of $B_1$.

For initial conditions in $B_1$, the contribution of the denominator in equation (\ref{eq:lyaptimemap}) is $\ln|\mathbf{w}_1|$
since $|\mathbf{w}_0|=1$.  For initial conditions (\ref{eq:b1initial})
with  initial vector $\mathbf{w}_0=[0,1]$ we can compute
$|\mathbf{w}_1|$ from equation (\ref{eq:thevs}) to obtain
\begin{equation}\label{eq:v1}
  \begin{array}{l} |\mathbf{w}_1|^2 = 9\alpha\left(\displaystyle\frac{(k-1/2)\pi}{2\alpha}
   \right)^{10/3} \\ 
   \;\;\;\;\;\;\;\;\;\;\;\;\;+  \left( 1 + 3b\alpha(-1)^k\left(\displaystyle\frac{(k-1/2)\pi}{2\alpha} \right)^{5/3} \right)^2. \end{array}
\end{equation}
For large $k$ we have
\begin{equation}
  |\mathbf{w}_1| \simeq (k\pi)^{5/3},
\end{equation}
and hence, for orbits of initial values (\ref{eq:b1initial}),
\begin{equation}\label{eq:denomforb1}
  \ln |\mathbf{w}_1| \simeq 5\ln(k\pi)/3.
\end{equation}
Substituting this result into (\ref{eq:lyaptimemap}), we find that the Lyapunov time 
for initial values on the boundary of $B_1$  behaves like,
\begin{equation}\label{eq:tlforb1}
  t_l \simeq \frac{3k\pi}{5\ln(k\pi)} \;\;\mbox{for large}\;\;
  k\in\mathbb{Z},
\end{equation}
where the survival time is determined by (\ref{eq:Tb1}). Since
$t_d\sim k\pi$ we can rewrite (\ref{eq:tlforb1}) as
\begin{equation}
  t_l \simeq \frac{3t_d}{5\ln(t_d)},
\end{equation}
which for large $t_d$ is of order $t_d/\ln(t_d)$.

\begin{figure}
  \includegraphics[width=\linewidth]{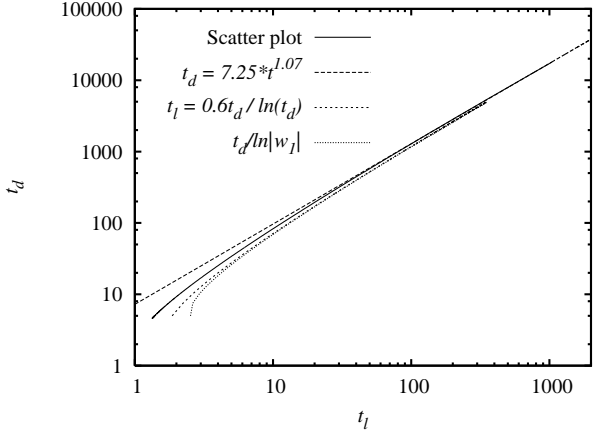}
  \caption{Various approximations to the scatter
    $(t_l,t_d)$ plot of orbits which escape after one excursion from
    the SOS.}
  \label{fig:rel1}
\end{figure}

Figure \ref{fig:rel1} shows various approximations to the scatter
$(t_l,t_d)$ plot (represented by the solid line) for orbits which
escape after one excursion from the SOS.  A best fit for a
curve in the form of the power law relationship (\ref{eq:pl}) is shown
by the long dashed line where $\beta=1.07$.  This curve does not fit
the scatter plot for small or large $t_l$ very well.  The dotted
curve represents the function $t_l=t_d/\ln|\mathbf{w}_1|$, for $|\mathbf{w}_1|$
given by (\ref{eq:v1}), which approximates
the scatter plot better than the power law relationship.  Finally, the
small dashed curve is the function $t_l=3t_d/(5\ln(t_d)$ which is
only a slightly better approximation to the scatter plot for small
$t_d$, but a much simpler one.

\subsection{A general functional relationship between $t_l$ and $t_d$}
\begin{figure}
  \includegraphics[width=\linewidth]{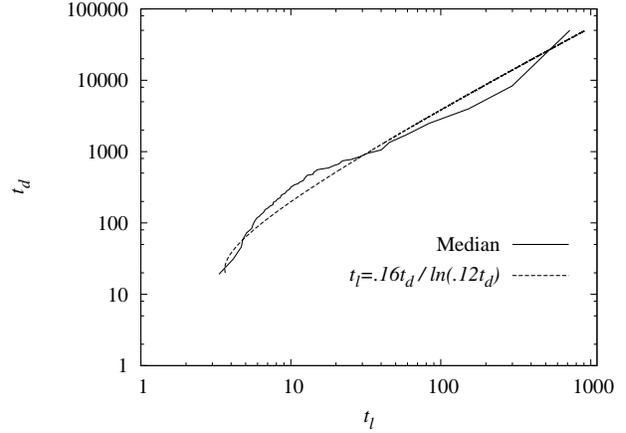}
  \caption{The solid line is the median curve in Figure
  \ref{fig:twopartmap} for $\varepsilon=0.61$.  The dashed line is the
  relationship (\ref{eq:genrelb}) with $\rho=0.16$ and $\nu=.12$.}
  \label{fig:newtlrel}
\end{figure}

Now consider the orbits which escape after two excursions from the
SOS. From equation (\ref{eq:lyaptimemap}) the Lyapunov time can be
approximated by,
\begin{equation}
  t_l \approx \frac{t_d}{\ln|\mathbf{w}_1| + \ln|\mathbf{w}_2|}.
\end{equation}
Following the result of the previous section, we assume that each term
in the denominator is proportional to the log of the time spent
between successive crossings of the SOS.  For escape after two
crossings these sum to $t_d$ and we take the two contributions to be
$\ln (ft_d)$ and $\ln ((1-f)t_d)$ where $0<f<1$, i.e.  

\begin{equation}\label{eq:genrela}
  t_l \sim \frac{t_d}{\ln \left(ft_d\right) + \ln\left((1-f)t_d\right)}.
\end{equation}
Re-arranging (\ref{eq:genrela}) we get
\begin{equation}
  t_l \sim \frac{t_d}{\ln\left(f\left(1-f\right)t_d^2\right)}.
\end{equation}
This relationship suggests fitting the data to curves which look like,
\begin{equation}\label{eq:genrelb}
  t_l = \rho\frac{t_d}{\ln\left( \nu t_d \right)}
\end{equation}
where $\rho$ and $\nu$ are constants.   In fact, adding successive
iterations produces a similar result so we take (\ref{eq:genrelb}) as
a general relationship and try to vary $\rho$ and $\nu$ to optimize a
fit to the data.  Figure \ref{fig:newtlrel} shows the median curve
(red solid curve) for 100000 initial values which escape within 1000
iterations of the map for $\varepsilon=0.61$.  We were able to approximately fit
the data with a function  like (\ref{eq:genrelb}) where $\rho=0.16$
and $\nu=.12$.

\section{Distributions}\label{sec:distributions}
Now we look at the distributions of the various quantities studied so
far and compare them to the results for the general 3-body problem. In  
the numerical experiments for the general
3-body problem in \cite{mikkola2007}, the authors fit the marginal
distribution of $t_d$ (for large values of $t_d$) and large values of
the ratio $Z=t_d/t_l$ to exponential probability density functions,
finding that
\begin{eqnarray}
  \psi(t_d) & \approx &\alpha \exp(-\alpha t_d), \;\; \alpha = 1/250,
  \label{eq:MTtd}\\
  \psi(Z) & \approx & \beta \exp(-\beta Z), \;\; \beta = 1/45.
  \label{eq:MTZ}
\end{eqnarray}  
The probability density for $t_l$ in the general 3-body problem was
obtained numerically and a theoretical explanation was given in
\cite{mikkola2007} as follows.  The authors assume that $t_d$ and $Z$ are
independent variables and so the marginal probability density
function, $\psi(t_l)$, can be determined via 
\begin{equation}\label{eq:MTtl}
	\begin{array}{rcl}
		\psi(t_l) & \approx & \displaystyle \int  \delta (t_l-t_d/Z)\psi(t_d)\psi(Z) dt_d dZ, \\
		& = & \alpha\beta/(\alpha t_l + \beta)^2.
	\end{array}
\end{equation} 
Their numerical results do not quite match up with (\ref{eq:MTtl}),
but the authors did fit their data with a function proportional to
$t_l^{-2}$.  The problem may be that  equation (\ref{eq:MTtl})
assumes that $t_d$ and $Z$ are independent variables.   This does
not seem to be likely considering Figure \ref{fig:scatter1}.  It may
be that, while the distributions provide satisfactory fits, an
expression like (\ref{eq:MTtl})  may not always be true.
\begin{figure}
  \includegraphics[width=\linewidth]{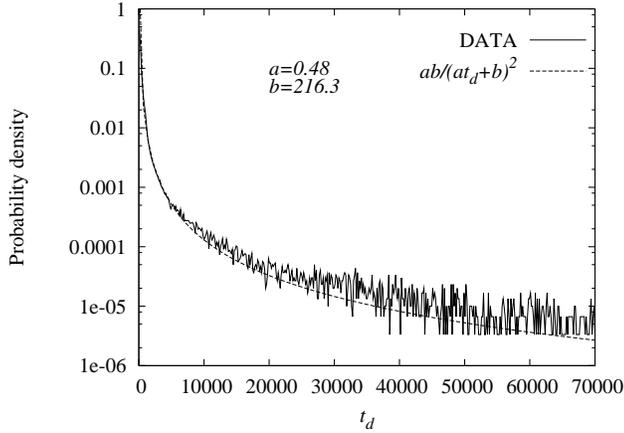}
  \caption{The solid curve is the probability density of $t_d$ for the
  numerical results with $\varepsilon=0.61$.  The dashed curve was found to be
  the best approximation of the stated form to the numerical results.}
  \label{fig:TDfit}
\end{figure}
\begin{figure}
  \includegraphics[width=\linewidth]{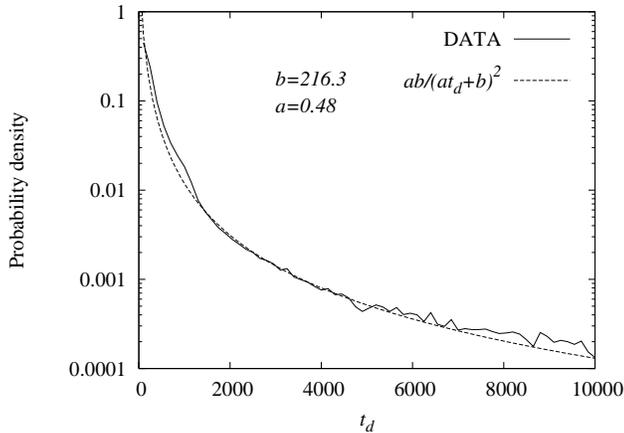}
  \caption{The solid curve is the probability density of $t_d$ for the
  numerical results with $\varepsilon=0.61$ for small $t_d$. The dashed curve
  was found to be the best approximation of the stated form to the
  numerical results.}
  \label{fig:tdSMALL}
\end{figure}

\begin{figure}
  \includegraphics[width=\linewidth]{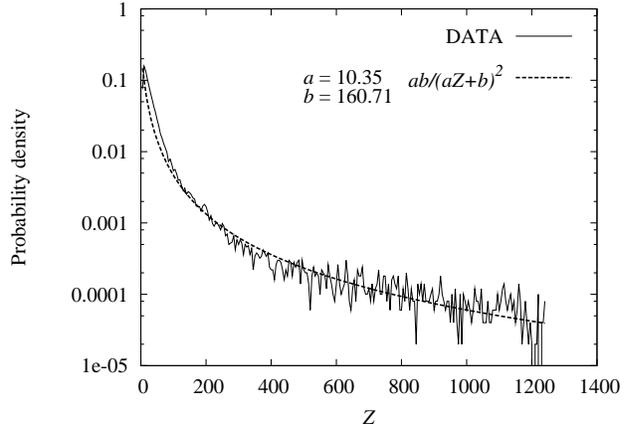}
  \caption{The solid curve is the probability density of $Z$ for the
  numerical results with $\varepsilon=0.61$. The dashed curve was found to be
  the best approximation of the stated form to the numerical results.}
  \label{fig:Zfit}
\end{figure}

Orbits computed with the map $\Phi$ also possess a similar power law
relationship between $t_d$ and $t_l$, for large enough eccentricities,
to those found in the general 3-body problem and the Sitnikov problem.
The map may
provide some insights into the distributions for $t_d$, $Z$ and $t_l$.
Figure \ref{fig:TDfit} shows the numerical results for the
distribution of $t_d$ values for the map $\Phi$ where $\varepsilon=0.61$.  The $t_d$
values are binned into intervals of length 200 along the whole range of
$0\leq t_d \leq 70000$ and the number of $t_d$
values in each bin is counted and divided by the total number of
initial conditions. 
It was found that a probability density function
\begin{equation}\label{eq:TDfit}
	\Psi(t_d) \approx a_1b_1/(a_1t_d+b_1)^2,
\end{equation}
where $a_1=0.48$ and $b_1=216.3$ best represents the data in Figure
\ref{fig:TDfit}. Figure \ref{fig:tdSMALL} shows the results for
 $t_d<1000$ which demonstrates that this is a good approximation for
small $t_d$.  Similarly, it was found
 that a probability density function for $Z$ which fits the numerical 
data satisfactorily is given by
\begin{equation}\label{eq:Zfit}
	\Psi(Z) \approx a_2b_2/(a_2Z+b_2)^2, 
\end{equation}
where $a_2=10.35$ and $b_2=160.71$ as shown in Figure \ref{fig:Zfit}.

\begin{figure}
  \includegraphics[width=\linewidth]{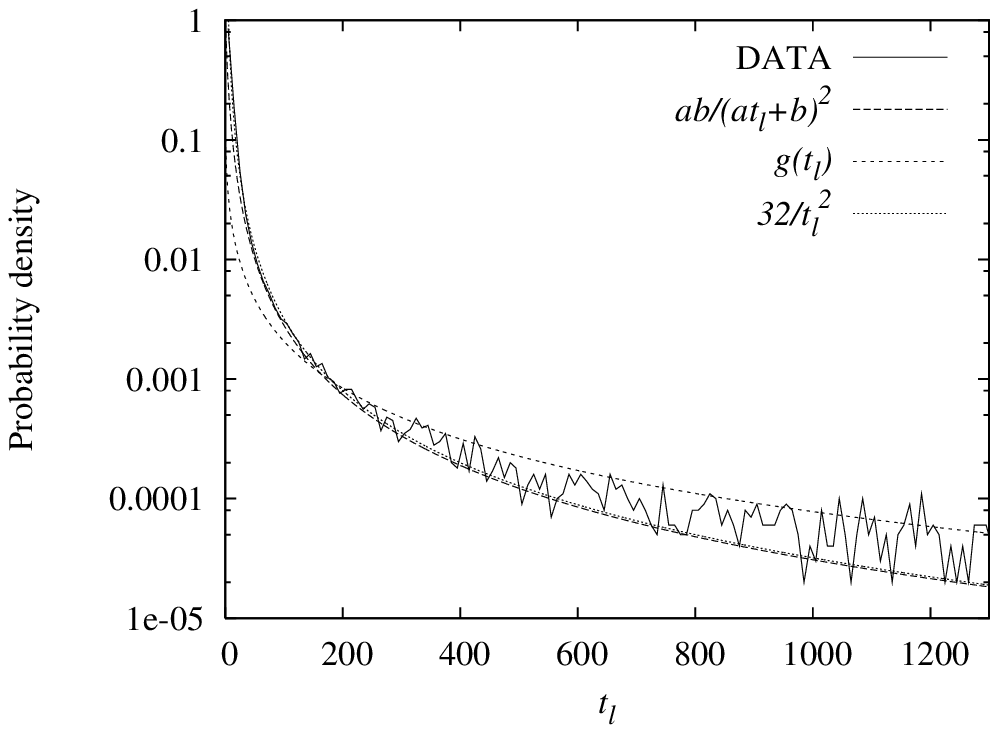}
  \caption{The solid curve shows the probability density of $t_l$ for
  the numerical experiments with $\varepsilon=0.61$.}
  \label{fig:tl}
\end{figure}
\begin{figure}
  \includegraphics[width=\linewidth]{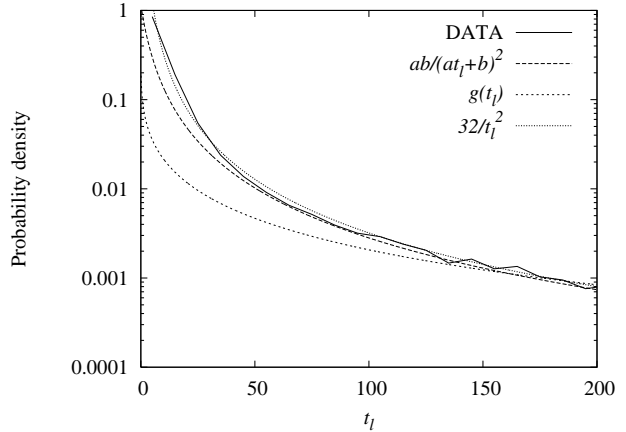}
  \caption{The solid curve shows the probability density of $t_l$ for
  the numerical experiments with $\varepsilon=0.61$ for small $t_l$.}
  \label{fig:tlSmall}
\end{figure}

The marginal probability density functions for $t_d$ and $Z$ are quite different to those for for the general 3-body
problem.  This means that the predicted density function
(\ref{eq:MTtl}) may not necessarily hold for  the
map.  Proceeding as in (\ref{eq:MTtl}), assuming the variables $t_d$
and $Z$ are independent, the density function for $t_l$ can be
determined by,
\begin{equation}\label{eq:TLpredict}
\begin{array}{rcl}
\Psi(t_l) & = &\displaystyle  \int \delta(t_l-t_d/Z)\Psi(t_d)\Psi(t_l)dt_d\;dt_l \\
	   & =  & -\displaystyle\frac{2b_1a_2\ln\left(\displaystyle \frac{b_1}{a_1t_l}\right)}{(-a_2b_1+b_2a_1t_l)^3} - \frac{1+\ln\left( \displaystyle \frac{b_1}{a_1t_l}\right)}{(-a_2b_1+b_2a_1t_l)^2} \\
	   & & +\displaystyle\frac{2a_1b_2t_l\ln\left(\displaystyle\frac{b_2}{a_2}\right)}{(-a_2b_1+b_2a_1t_l)^3}-\displaystyle\frac{1+\ln\left(\displaystyle\frac{b_2}{a_2}\right)}{(-a_2b_1+b_2a_1t_l)^2}.
\end{array}	   
\end{equation}
Figure \ref{fig:tl} shows the numerical results for the map $\Phi$ for
$\varepsilon=0.61$.  The solid curve shows the data from the experiments, and the
medium dashed curve is the predicted curve (\ref{eq:TLpredict}) where
$g(t_l)=\Psi(t_l)$.  Note that the curve does not quite fit the data
from the experiments for small $t_l$ (Figure \ref{fig:tlSmall}).  The
data was then fitted to a curve of the form,
\begin{equation}\label{eq:TLfit}
	\Psi(t_l) \approx ab/(at_d+b)^2,  \;\; \mbox{where} \;\; a=0.53,\; b=2.75,
\end{equation}
shown by the short dashed curve.  Again, this failed to fit the data for
small $t_l$.  It was noted that the data is better approximated by the
curve,
\begin{equation}
	f(t_l)=32/t_l^2.
\end{equation}
Interestingly, this is a similar function to that which fits the numerical
results for the general 3-body problem.
 This  suggests that the probability distribution associated with the
 Lyapunov time
may not necessarily be the result of the probability distribution of
survival
times and the probability distribution of the variable $Z$.

\section{Conclusions}

One of the results of this investigation is demonstrating a
relationship between the Lyapunov time and the survival time for the
Sitnikov problem which is similar to that for the general 3-body
problem found in \cite{mikkola2007}.   This is surprising as the
Sitnikov problem
is rather different compared to the 3-body systems discussed in \cite{mikkola2007}.  With the use of an
approximate Poincar\'e map we were able to delineate regions of escape
on a surface of section so as to construct initial conditions for the map.
By studying the relationship between the Lyapunov time and the survival
time with initial conditions in distinct escape regions, we were able
to analytically obtain  a new functional relationship between $t_l$ and $t_d$ given by
$t_l=\rho t_d/\ln (\nu t_d)$ where $\rho$ and $\nu$ are constants. As the
$(t_l,t_d)$ 
scatter plots for the Sitnikov problem are similar to the $(t_l,t_d)$
scatter plots for the
general 3-body problem, we conjecture that the new functional relationship
between $t_l$ and $t_d$ presented above may also be valid for the general 3-body problem.

Interestingly, the marginal distributions for the quantities $t_d$ and
$Z$ were found to be  different than   for the general 3-body
problem. The reason for this is not known although it may just be that
in this study we considered the entirety of the numerical results and
not just large values.  Treating  $t_d$ and $Z$ as independent
variables, we were able to derive a marginal distribution for $t_l$
which did not quite capture the numerical results.  It seems unlikely
though that $t_d$ and $Z$ are independent quantities which may be why
the theoretical predicted distribution $\Psi(t_l)$ did not quite match
the numerical results.   Interestingly, as for the general 3-body problem, it 
was found that a function proportional to $t_l^{-2}$ fits the numerical
distribution of $t_l$ well.  This may point to a more general property
which may be valid for 3-body problems which experience large
perturbations to an escaping mass.

\label{lastpage}

\end{document}